\documentclass[10pt,twocolumn]{article}

\usepackage[a4paper,margin=2cm]{geometry}
\usepackage[T1]{fontenc}
\usepackage[utf8]{inputenc}
\usepackage{newtxtext,newtxmath}

\usepackage{amsmath}
\usepackage{graphicx}
\usepackage{booktabs}
\usepackage{multirow}
\usepackage{array}
\usepackage{xcolor}
\usepackage{hyperref}
\usepackage[numbers]{natbib}
\usepackage{url}

\usepackage{caption}
\usepackage{setspace}

\usepackage{titlesec} 
\titleformat{\section} 
{\normalsize\bfseries} 
{\thesection} {0.5em} 
{} 

\titleformat{\subsection} 
{\small\bfseries}
{\thesubsection}
{0.5em} 
{} 

\renewenvironment{abstract} 
{ \small\bfseries
\noindent\textbf{Abstract} } {}

\date{}

\usepackage{etoolbox}

\makeatletter
\patchcmd{\thebibliography}
  {\list{\@biblabel{\@arabic\c@enumiv}}}
  {\list{\@biblabel{\@arabic\c@enumiv}}%
   {\setlength{\itemsep}{0pt}%
    \setlength{\parskip}{0pt}%
    \setlength{\parsep}{0pt}}}
  {}{}
\makeatother

\begin{document}

\twocolumn[
\begin{center}
\vspace*{-0.5cm}

{\LARGE
OPC UA Shared-Memory: Conceptual Elaboration and Prototypical Implementation Using Iceoryx2
\par}

\vspace{1.5em}
{
Thomas Barth$^{1}$ \quad
Tatjana Legler$^{1,2}$ \quad
Achim Wagner$^{1,2}$ \quad
Martin Ruskowski$^{1,2}$
\par}

\vspace{0.5em}

{\small
$^{1}$ Chair of Machine Tools and Control Systems, RPTU Kaiserslautern-Landau,\\
Gottlieb-Daimler-Straße 42, 67663 Kaiserslautern, Germany\\[0.2em]
$^{2}$ Innovative Factory Systems, German Research Center for Artificial Intelligence,\\
Trippstadter Straße 122, 67663 Kaiserslautern, Germany
\par}

\vspace{1em}
\end{center}
]

\begin{spacing}{0.9}
\begin{abstract}
- The increasing virtualization of automation software leads to a growing co-location of heterogeneous applications on shared edge devices and fosters modern microservice architectures. As a result, traditional OT software and modern IT-driven higher-level automation applications are increasingly executed on the same host system, enabling improved synergy and more efficient interaction between these domains. This development imposes strict requirements on the underlying communication infrastructure, particularly regarding latency and determinism. At the same time, there is a lack of middleware solutions that address these requirements while incorporating modern concepts such as standardized semantic information models.
Building on prior work, this paper further develops a previously proposed concept of extending the OPC UA communication model for microservice-based applications and suggests the use of shared-memory as a transport mechanism for optimized intra-host communication. In addition, a first prototypical implementation based on a modified OPC UA library using Iceoryx2 is presented and evaluated in a proof-of-concept study by measuring function call latencies on the user level and comparing them to the original TCP-based OPC UA implementation. The results show consistently lower latencies for the shared-memory approach, while also indicating remaining optimization potential due to current implementation constraints.
\end{abstract}
\end{spacing}

\vspace{0.5em}
\noindent\textbf{Keywords:} OPC UA, Microservices, Shared-Memory, Virtualization

\section{Introduction}
Industrial machines and production systems must become more and more flexible, resilient, and intelligent in order to meet the demands of modern manufacturing environments \cite{ElMaraghy2005-ak}. This development is accompanied by growing requirements regarding the software capabilities of machines. The functional scope is expanding, the number of digital software and hardware components increasing, and adaptations must be implemented within shorter timeframes due to rapidly changing production requirements. This development, combined with the increasing amount of digital information, results in more complex decision criteria and, consequently, more sophisticated control logic.
In many machines, the programmable logic controller (PLC) remains one of the most important software components. Historically developed as a digital replacement for hardwired logic circuits between sensors and actuators, PLCs have evolved over time but remain particularly optimized for hardware-oriented and deterministic control tasks. With the emergence of Industry 4.0 concepts and the resulting shift towards increasingly software-defined production systems, the realization of these concepts using traditional PLC technology is reaching its conceptual limits \cite{Sunilkumar_Kollamolel2025-ak}.

As a consequence, required software solutions progressively emerge outside the classical operational technology (OT) domain. Beyond PLCs, industrial automation is incorporating modern software components, general-purpose programming languages and development tools from the information technology IT sector, where design paradigms are shaped by software- and data-driven problem formulations. Therefore, automation software evolves into a composition of heterogeneous software components operating at different abstraction levels and following distinct development approaches \cite{Kok2024-ip}.

The interaction between these software domains imposes strict requirements on the underlying communication infrastructure, as data exchange in automation must be reliable, fast and meet tight timing constraints. At the same time, PLC systems are increasingly being virtualized, particularly through containerization \cite{Goldschmidt2016-nn}, leading to their decoupling from dedicated control hardware. As a result, traditional OT software and automation applications influenced by modern software engineering paradigms are increasingly co-located on shared edge devices. This co-location creates new potential for a more efficient and consistent interconnection between PLCs and higher-level application software.

Although this potential has been recognized in the market, existing middleware solutions are typically proprietary and do not combine an open and standardized semantic meta-model with deterministic capabilities. To address this gap, an initial concept for an intra-host middleware based on OPC Unified Architecture (OPC UA) was proposed in \cite{barth_towards_2025}, extending the communication model of the standard and using shared-memory as the underlying transport mechanism.

While the initial publication introduced the general idea, this paper refines the concept and presents a first prototype based on the shared-memory framework Iceoryx2 \cite{iceoryx2_github}. In addition, a proof-of-concept evaluation is conducted to compare service reaction times with a network-based implementation.

\section{Background and Related Work}
This section provides an overview of the relevant publications in the literature on the topics of OPC UA, microservices in automation and the use of shared-memory for efficient intra-host communication.

\subsection{OPC UA}
OPC UA is a platform- and vendor-independent communication standard that combines service-oriented communication with standardized information modeling \cite{opc_foundation_part1}. It is widely used in automation across a broad range of applications, from machine monitoring and data acquisition \cite{Hastbacka2014-fk} to the semantic description and control of cyber-physical production systems \cite{Muller2017-ro}. To address evolving requirements, OPC UA has been continuously extended, most notably by the introduction of the Publish/Subscribe (PubSub) communication model. This extension enables scalable and efficient data distribution, particularly for high-frequency and many-to-many communication scenarios, and complements the traditional Client/Server model \cite{opc_foundation_Part14PubSub}. While the PubSub model improves scalability and reduces communication overhead in distributed environments, it only supplements the Client/Server model and does not provide a request-response pattern \cite{opc_foundation_Part14PubSub} required by many service-oriented automation use cases. In such architectures and particularly in microservice-based systems, an additional hurdle appears, as the growing number of distributed services introduces challenges in managing and organizing large amounts of data and semantic models.

In this context, Tripathy et al. \cite{tripathy_opc_2022} address an aspect of this challenge, namely that OPC UA Server discovery alone does not allow consumers to dynamically identify and bind relevant services at runtime, based on service attributes. Although OPC UA provides mechanisms for Server discovery and browsing, consumers still need to determine which Server and which nodes are relevant, which increases engineering effort and limits flexibility in dynamic environments. To address this, the authors integrate OPC UA with the Eclipse Arrowhead Framework, introducing a centralized service registry and orchestration mechanism that enables service discovery and binding based on metadata at runtime.

However, both the application of OPC UA as a communication mechanism in microservice-based automation systems and its native evolution towards supporting such scenarios, have not yet been systematically investigated.

\subsection{Microservices in Automation}
Microservice-based architectures, originating from the IT domain, are increasingly being adopted in industrial automation. Compared to monolithic and traditional service-oriented designs, they enable a more flexible and modular structuring of software components and services, allowing individual components to be developed, deployed, and operated more independently \cite{De_Lauretis2019-dj}. This is further supported by advances in virtualization and containerization, which enable the deployment and execution of such components on shared hardware.

Sarkar et al. \cite{Sarkar2018-jp} provide practical insights into the transformation of an existing industrial automation system from a monolithic to a microservice-based architecture using container technologies. The authors highlight that these approaches are gaining relevance in industrial environments, particularly due to increasing virtualization and containerization, which enable flexible scalability, improved resource utilization, and increased adaptability. At the same time, they show that migrating existing systems to such architectures is non-trivial and requires significant structural adaptations, particularly due to existing strong dependencies between components.

Bigheti et al. \cite{Bigheti2019-aq} investigate the use of microservice-oriented architectures in industrial automation and argue that they facilitate flexible integration of automation and IT systems. Microservices are thereby considered an evolution of service-oriented architectures, aiming to increase modularity and enable independently deployable functional units. Through a “Control-as-a-Service” approach, the authors demonstrate that even classical automation functionalities, such as closed-loop control, can in principle be realized as distributed services.

However, the measurements of Bigheti et al. reveal important limitations. In particular, the experimental evaluation shows significant communication latency and jitter, resulting in non-deterministic system behavior. This indicates that while microservice architectures provide high flexibility and interoperability, they impose strict requirements on the underlying communication infrastructure in order to support time-critical automation tasks.

Overall, existing work indicates a growing interest in applying microservice-based principles in industrial automation. However, there is still limited focus on communication or middleware approaches specifically tailored to the requirements of such systems. As demonstrated by \cite{Bigheti2019-aq}, typical automation applications impose strict constraints regarding latency, determinism, and reliability, which are only partially considered by existing solutions.

\subsection{Shared-Memory Communication}

While virtualization and containerization are increasingly adopted in industrial automation \cite{Goldschmidt2016-nn, Queiroz2023}, the efficient realization of low-latency intra-host communication between \mbox{co-located} software components remains insufficiently explored in this context. Related challenges have, however, been addressed more extensively in other domains, particularly within the IT domain and in robotics.

In the context of cloud-native microservice applications, Khasgiwale et al. \cite{Khasgiwale2023-rj} analyze inter-container communication and show that network-based communication with a conventional message broker between co-located containers introduces significant latency and resource overhead, which can limit the efficient execution of such applications on local edge systems. To address this, they propose a shared-memory-based approach that enables direct data exchange between containers on the same host. Their evaluation demonstrates substantial performance improvements. At 10\,kB, the shared-memory-based approach achieves latencies of approximately 13\,µs compared to about 251\,µs for \mbox{RabbitMQ}. At 10\,MB, latencies increase only slightly to around 18\,µs, whereas \mbox{RabbitMQ} reaches up to 85.67\,ms. In addition, CPU and memory consumption in the shared-memory measurements are reduced.

Similar challenges and solutions have been identified in robotics middleware. Wang et al. \cite{Wang2019-po} indicates that serialization and multiple memory copies in socket-based communication are major sources of latency in frameworks such as the Robot Operating System (ROS) and ROS2, particularly for large messages and multiple subscribers. To improve this, the authors propose a hybrid communication approach in which control information is transmitted via sockets, while data-intensive payloads are exchanged via shared-memory, enabling a zero-copy mechanism. Their evaluation shows that, for message sizes of 4\,MB, the latency in ROS decreases from tens of milliseconds to hundreds of microseconds, while in ROS2 it drops from hundreds of milliseconds to below one millisecond.

These findings highlight the potential of shared-memory for efficient intra-host communication in modular and microservice-based systems. However, comparable investigations in co-located industrial automation applications remain limited and the integration with standardized semantic communication models has not yet been addressed.

\section{Conceptual Foundation}
To enable a service-oriented microservice architecture using OPC UA as middleware, conceptual challenges of the classical Client/Server model were discussed in \cite{barth_towards_2025}. To describe the participating applications, the roles of \emph{Provider} and \emph{Consumer} were introduced. A \emph{Provider} offers data or methods, while a \emph{Consumer} accesses them. An application can assume one or both roles simultaneously.

When applying the classical OPC UA Client/Server model within a microservice architecture, each application acting as a \emph{Provider} implements its own OPC UA Server, and each application acting as a \emph{Consumer} implements at least one OPC UA Client. As the number of applications increases, complexity grows, particularly in the discovery and browsing processes, since no central directory of available data exists. 

To improve transparency of available data and enhance the scalability of the communication model, a central instance, the \emph{DataDirectoryServer} (DaD-Server), was proposed. This Server is implemented as a conventional OPC UA Server, in which all \emph{Providers} register their exposed data. The \mbox{\emph{DaD-Server}} aggregates this data within its address space and thus provides a central directory. As a result, the discovery process is significantly simplified and can be performed using a single client-connection per \emph{Consumer}.

In \cite{barth_towards_2025}, the \emph{DaD-Server} was intended not only to manage data registration but also to coordinate access and act as a broker between \emph{Consumers} and \emph{Providers}. This approach was revised in the prototype presented in this paper. Forwarding every individual data request would require continuous involvement of the \emph{DaD-Server} in data and service communication and would introduce an additional communication step. This results in increased latency and a stronger dependency on a central instance, thereby raising the risk of a single point of failure. For this reason, a peer-to-peer approach between \emph{Consumers} and \emph{Providers} is adopted in the following.

\section{Architectural Refinement and System Design}
For the development of a first prototype, this section specifies and revises key architectural aspects of the approach to establish a foundation for a concrete implementation. These include the separation of information provisioning between the \emph{DaD-Server} (later renamed to \emph{SemanticRegistryServer}) and \emph{Providers}, the registration and modeling of provided data, and the access to semantic information, data values, and methods by a \emph{Consumer}.

\subsection{Centralized Modeling of Semantic Information}
In a microservice architecture with multiple \emph{Providers}, the classical OPC UA Client/Server approach leads not only to limited data transparency, but also to redundant representations of information models, particularly of the base~model, across multiple Server instances. While this does not introduce functional limitations, it creates an additional structural complexity of the overall architecture. To reduce this, an explicit separation of information provisioning between the \emph{DaD-Server} and \emph{Providers} is defined.

If the overview of all available data is provided through the \emph{DaD-Server}, it is consistent to also consolidate the corresponding information models at this location. \emph{Consumers} can therefore not only discover all available data points through a single client connection, but also access their associated semantic information.

\emph{Providers}, as owners of the data sources, expose only the respective data values via their server-interface, without redundantly describing the associated structures or underlying model. Consequently, responsibility for the model, including type definitions and structural descriptions, is separated from the value-providing communication. The semantic description of individual data points is therefore managed centrally. This principle is applied analogously to methods.

Reflecting this further refinement and the more precisely defined role of the \emph{DataDirectoryServer}, the component is renamed to \emph{SemanticRegistryServer} (SR-Server) in order to better describe its function as a centralized repository for semantic information and a registration point for providers and their resources. While the \emph{SR-Server} ensures a consistent semantic description and overall system transparency, \emph{Providers} focus on the efficient and performant provisioning of operational data values and functions.

\subsection{Registration and Modeling of Provided Data}
When a \emph{Provider} exposes data, these must be registered on the \emph{SR-Server} according to the proposed concept. For registration, the \emph{Provider} connects to the server-interface of the \emph{SR-Server} via an OPC UA Client. The available data is then added and modeled in the address space of the central Server using the OPC UA service \emph{AddNodes}. The \emph{NodeId} assigned on the \emph{SR-Server} is stored by the \emph{Provider} and later used for access to the corresponding resource through the \emph{Providers} own server-interface.

\begin{figure}[htbp]
  \centering
  \includegraphics[width=70mm]{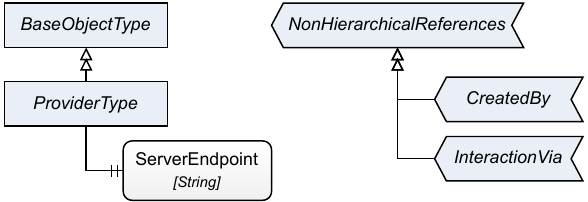}
  \caption{Added object type and references}
  \label{fig:NewObjectAndRefs}
\end{figure}

To uniquely identify both \emph{Providers} and their registered data in the address space, they are explicitly modeled and referenced on the \emph{SR-Server}. For this purpose two new modeling elements are introduced: the object type \emph{ProviderType} with the property \emph{ServerEndpoint}, and the references \emph{CreatedBy} and \emph{InteractionVia}, see Fig.~\ref{fig:NewObjectAndRefs}.

\begin{figure}[htbp]
  \centering
  \includegraphics[width=71mm]{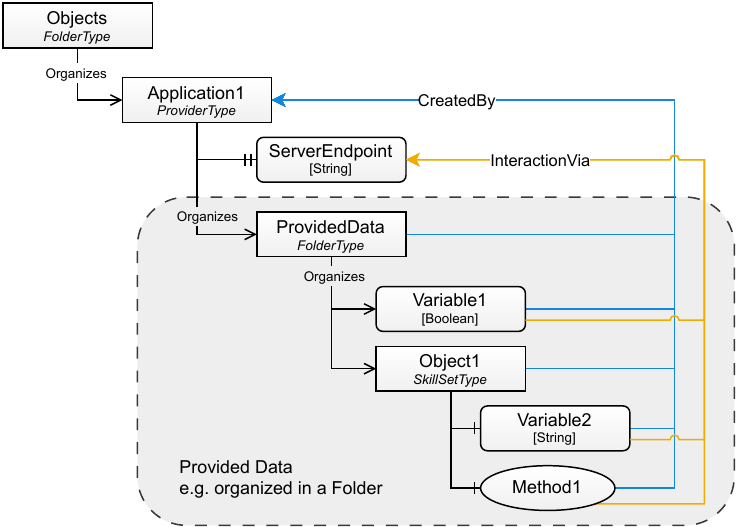}
  \caption{Example of a \emph{Provider} and the representation of its data on the \emph{SR-Server}}
  \label{fig:ExampleProvidedData}
\end{figure}

The \emph{ProviderType} represents an instance of a \emph{Provider} and is created in the Objects folder in the \emph{SR-Servers} address space. The property \emph{ServerEndpoint} describes the access point to the \emph{Providers} server-interface via a specific endpoint address. Using the \emph{CreatedBy} reference, all nodes created by a \emph{Provider} on the \emph{SR-Server} are linked to the corresponding \emph{Provider} instance. Nodes such as variables or methods, whose values or invocation are provided directly by the \emph{Provider}, are additionally annotated with the reference \emph{InteractionVia}, which points to the corresponding server-endpoint. In Fig.~\ref{fig:ExampleProvidedData}, an exemplary modeling of a \emph{Provider} and its data is shown.

\subsection{Provisioning and Consumption of Registered Data}
To interact with data provided by a \emph{Provider}, an application acting as a \emph{Consumer} first connects to the \emph{SR-Server}. By browsing the address space, the \emph{Consumer} can identify all registered data points as well as their corresponding \emph{Providers}. Semantic information for the registered data can be directly retrieved from the \emph{SR-Server}.

To access the data value of a variable or call a method associated with the reference \emph{InteractionVia}, the \emph{Consumer} must establish an additional client-connection to the respective \emph{Provider}. For this purpose, the \emph{Consumer} connects via the referenced endpoint address to the \emph{Providers} server-interface. Data access then occurs directly between \emph{Consumer} and \emph{Provider}.

\begin{figure}[htbp]
  \centering
  \includegraphics[width=\linewidth]{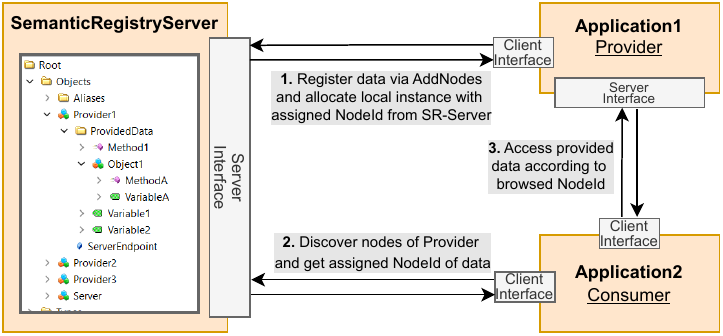}
  \caption{Interaction between \emph{Provider}, \emph{SR-Server} and \emph{Consumer}}
  \label{fig:SystemInteraction}
\end{figure}

The server-interface of a \emph{Provider} is implemented as a reduced variant of an standard OPC UA Server. It only provides those OPC UA services required for connection establishment as well as for reading and writing data points and calling methods. The address space consists solely of a mapping between the \emph{NodeId} and the corresponding data or function source. Using the \emph{NodeIds} retrieved from the \emph{SR-Server}, a \emph{Consumer} can read the corresponding data value directly from the \emph{Provider} using the OPC UA service Read. Fig.~\ref{fig:SystemInteraction} provides an overview of the step-by-step interaction between a \emph{Provider} and the \emph{SR-Server} during data registration, as well as data discovery and access by a \emph{Consumer}.

\section{Implementation}
The goal of the implementation is to realize a prototype of the proposed architecture and to enable a subsequent evaluation of communication latency. The functional scope is deliberately limited to core OPC UA functionalities. In particular, browsing of the address space as well as basic read and write operations are implemented. More advanced aspects of the OPC UA standard, such as security mechanisms or sessions, are not considered within the prototype.

The following sections first describe the selection of a suitable shared-memory framework, followed by the implementation design of the prototype, and subsequently discuss an identified challenge during implementation and the resulting limitations.

\subsection{Technology Selection for Shared-Memory Communication}
As discussed in \cite{barth_towards_2025}, communication via shared-memory represents a complex challenge, especially when high reliability and deterministic communication behavior is required.

For this reason, a suitable interprocess communication (IPC) library will be used for the implementation, which abstracts the underlying operating system mechanisms as well as access to and management of the corresponding memory regions through a user-friendly interface. Since an open and freely available solution is intended, only open source libraries are considered. Furthermore, the implementation will be realized in C++, as it provides the required performance to enable a meaningful comparison with high-performance OPC UA implementations.

During the evaluation, several open source libraries supporting shared-memory communication were identified. Many of these solutions originate from the IT domain and are designed for efficient and low latency communication of large data volumes between applications in Server or cloud environments, for example shmipc \cite{shmipc_go_github} or Flow-IPC \cite{flow_ipc_github}. However, an explicit focus on deterministic communication behavior or usage in real time systems is not emphasized, which limits their suitability for the requirements considered in this work.

In addition to these libraries, the established middleware Data Distribution Service (DDS), specifically its open source implementation Fast DDS, also supports shared-memory-based communication. In DDS, however, this mechanism represents more of an optional transport optimization within a broader middleware stack \cite{fastdds} rather than the primary communication mechanism. For the approach pursued in this work, a specialized IPC library therefore appears more suitable.

Another frequently referenced C++ library in the context of shared-memory communication is Boost \cite{boost}. However, Boost primarily provides low level building blocks, meaning that the actual implementation of a communication model must largely be realized by the application itself.
In addition, several smaller libraries such as cpp-ipc \cite{cpp_ipc_github} or shadesmar \cite{shadesmar_github} exist. These projects are often maintained by individual developers and partly exhibit limited documentation and functionality, which makes them unsuitable for the intended use.

Among the available solutions, the framework Iceoryx2 (Iox2) \cite{iceoryx2_github} was identified as a feasible option. Iceoryx was originally developed in the context of safety critical real time systems in autonomous driving and is specifically designed for high-performance and reliable interprocess communication via shared-memory. Iox2 follows a zero copy approach, in which data is transferred between processes without additional copying. As a result, transmission time is largely independent of payload size, which provides performance benefits and supports deterministic communication behavior. Furthermore, Iox2 offers different communication models, such as Publish/Subscribe and Request/Response. The latter is particularly suitable for implementing the service oriented OPC UA communication model. In addition to its native Rust implementation, bindings for multiple programming languages, including C, C++, Python, and C\# are available, which facilitates integration into heterogeneous software environments. Additionally, the project is actively maintained by a commercial company and supported by a comparatively large community. Based on these characteristics, Iox2 provides a suitable foundation for the implementation and is therefore selected for the prototype.

\subsection{Prototype Implementation with Iceoryx2}
Since implementing the OPC UA standard from scratch would require significant development effort, the C++ project FreeOpcUa \cite{freeopcua_cpp} was selected as the basis for the prototype. Although it implements only fundamental parts of the OPC UA specification, its comparatively simple and modular codebase makes it suitable for experimental modifications.

To realize communication via shared-memory, the original network transport layer of the codebase was replaced with an implementation based on Iox2. For this purpose, the Request/Response communication model was used. In this model, an application defines an Iox2 service with specified request and response data types, which can be called by another application acting as a Client. Each service is assigned a unique name through which it can be identified within the Iox2 namespace. Data transmission is performed entirely via shared-memory.

To implement a server-interface, the OPC UA \emph{Service Sets} relevant for the prototype, such as \emph{Views}, \emph{Attributes}, and basic \emph{NodeManagement} services, were mapped to corresponding Iox2 services including their respective request and response data types. A uniform naming convention was introduced for structured addressing. Each service follows the scheme {\small\texttt{\textless ApplicationName\textgreater/\textless ServiceSet\textgreater/\textless Service\textgreater}}, for example {\small\texttt{DataDirectoryServer/Attributes/Read}}. This convention allows all services of an instance to be addressed via a common prefix within the Iox2 namespace.

For the client-interface, a corresponding counterpart consisting of Iox2 service-clients was implemented. These directly connect to and interact with the available services of a server-interface.

Based on these Iox2 interfaces, the \emph{SR-Server}, \emph{Providers}, and \emph{Consumers} were implemented. The access point of a \emph{Provider} is described on the \emph{SR-Server} via an endpoint of the form {\small\texttt{opc.iox2://\textless ApplicationName\textgreater}}. From this instance identifier, the corresponding Iox2 service addresses can be derived according to the defined naming convention. \emph{Consumers} can instantiate additional client-interfaces at runtime to access data provided by specific \emph{Providers}. The address space of a \emph{Provider} is implemented as a simple map between registered \emph{NodeIds} and instances of the corresponding data points.

\subsection{Data Type Limitations}
Due to the zero copy approach of Iox2, transmitted data types must be fully representable in shared-memory. Dynamically growing types that rely on heap memory, such as strings or vectors from the C++ standard library, as well as constructs based on pointers or references, cannot be used directly \cite{iceoryx_restrictions}. This presents a challenge, particularly for the transmission of complex OPC UA data types. Although Iox2 provides stack-based implementations of common data types with fixed sizes, adapting the existing codebase completely to these types would require substantial modifications and involve significant implementation efforts.

As a pragmatic temporary workaround, the existing OPC UA serialization of the network stack is currently used. Data is serialized before transmission and transferred as a byte buffer via Iox2. Although this approach introduces additional copy operations during sending and receiving and therefore deviates from the zero copy principle, improved performance compared to the original network-based communication is still expected.

\section{Performance Evaluation}
This section evaluates the proposed approach and its implementation with respect to general feasibility and communication latency. The focus is on a proof of concept and an initial comparison between the shared-memory-based modification and the original network-based OPC UA library. The evaluation deliberately considers only the reactivity of service oriented data transmission and not the behavior of the complete system architecture. The \emph{SR-Server} is not included, since it is not part of the performance critical data path.

\subsection{Setup}
The measurements were conducted on a Lenovo~ThinkCenter~m80t~Gen~3 equipped with an Intel~i5~12400 processor. As operating system a natively installed Ubuntu 24.04.3~LTS was used. All applications were executed directly on the system without virtualization and measurements refer exclusively to intra-host communication between processes on the same system.

\subsection{Methodology}
To evaluate communication latencies, function calls at application level are analyzed. For this purpose, the high level API functions \emph{GetValue()} and \emph{SetValue()} of the C++ Class\,Node are used, which internally invoke the OPC UA attribute services \emph{Read} and \emph{Write}. The function execution includes the complete processing chain, from serialization and deserialization of the data, to transmission via the underlying communication mechanism, and access to the address space.

Measurements are performed for two data types, \emph{Uint32} with 4\,bytes and a vector consisting of 1024\,\emph{Uint32} elements with 4096\,bytes, in order to analyze the influence of payload size on the overall execution time. For each combination of data type and operation, three measurement series with 1000 iterations each are conducted to reduce random disturbances.
The measurements are performed between \emph{Consumer} and \emph{Provider} in the prototype and between Client and Server in the original codebase. No security mechanisms were enabled in the original codebase.

\subsection{Results and Discussion}

The measurement results for \emph{GetValue()} and \emph{SetValue()} operations are summarized in Table~\ref{tab:measurements_4bytes} and Table~\ref{tab:measurements_4096bytes} with all reported values rounded to the nearest integer using standard rounding.

%============================================================================
\begin{table}[h]
\centering
\scriptsize
\caption{Latency comparison (µs) of API function calls between Iox2 and TCP implementation with 4~bytes payload}
\begin{tabular}{ll|ccc|ccc}
\hline
\textbf{Operation} & \textbf{Metric}
 & \multicolumn{3}{c|}{\textbf{Iox2}}
 & \multicolumn{3}{c}{\textbf{TCP}} \\
 &  & \textbf{M1} & \textbf{M2} & \textbf{M3} & \textbf{M1} & \textbf{M2} & \textbf{M3} \\
\hline

\multirow{4}{*}{\emph{GetValue()}}
 & Min    & 15  & 17 & 17  & 31  & 30  & 31 \\
 & Median & 20  & 21 & 20  & 33  & 34  & 33 \\
 & Mean   & 31  & 35 & 36  & 39  & 40  & 42 \\
 & Max    & 2173 & 1787 & 1085 & 1001 & 1255 & 815 \\

\hline

\multirow{4}{*}{\emph{SetValue()}}
 & Min    & 15 & 17 & 16 & 31 & 31 & 30 \\
 & Median & 20 & 21 & 20 & 34 & 35 & 35 \\
 & Mean   & 34 & 37 & 39 & 45 & 48 & 44 \\
 & Max    & 1829 & 756 & 1573 & 1291 & 1186 & 1277 \\

\hline
\end{tabular}
\label{tab:measurements_4bytes}
\end{table}

%============================================================================
%============================================================================
\begin{table}[h]
\centering
\scriptsize
\caption{Latency comparison (µs) of API function calls between Iox2 and TCP implementation with 4096~bytes payload}
\begin{tabular}{ll|ccc|ccc}
\hline
\textbf{Operation} & \textbf{Metric} %Statistic
 & \multicolumn{3}{c|}{\textbf{Iox2}}
 & \multicolumn{3}{c}{\textbf{TCP}} \\
 &  & \textbf{M1} & \textbf{M2} & \textbf{M3} & \textbf{M1} & \textbf{M2} & \textbf{M3} \\
\hline

\multirow{4}{*}{\emph{GetValue()}}
 & Min    & 37 & 36 & 36 & 51 & 52 & 51 \\
 & Median & 40 & 41 & 41 & 56 & 55 & 56 \\
 & Mean   & 50 & 52 & 49 & 64 & 65 & 67 \\
 & Max    & 777 & 754 & 532 & 907 & 1051 & 1201 \\

\hline

\multirow{4}{*}{\emph{SetValue()}}
 & Min    & 38 & 38 & 36 & 53 & 53 & 52 \\
 & Median & 41 & 42 & 41 & 57 & 57 & 57 \\
 & Mean   & 57 & 56 & 55 & 66 & 66 & 70 \\
 & Max    & 948 & 961 & 1165 & 952 & 830 & 1621 \\

\hline
\end{tabular}
\label{tab:measurements_4096bytes}
\end{table}
%============================================================================

For small payloads of 4\,bytes, the median latencies of the shared-memory implementation are approximately 20\,-\,21\,µs, while the network-based implementation reaches approximately 33\,-\,35\,µs. For larger payloads of 4096\,bytes, latencies increase in both cases but remain consistently lower for the shared-memory variant, at approximately 40\,-\,42\,µs compared to 55\,-\,57\,µs.

No significant differences are observed between \emph{GetValue()} and \emph{SetValue()}. Since the transmitted payload is similar in both cases, this indicates that processing times for read and write access to the address space do not introduce notable differences.

The minimum values show similar behavior and can be interpreted as an approximation of the best case. Here as well, an advantage for the shared-memory-based communication is observed, with a consistent latency reduction of about 15\,µs. This corresponds to an improvement of 50\,\% for small payloads and around 30\,\% for larger payloads (related to best-case minimum latencies).

It is notable that the influence of payload size is comparable in both implementations. This shows that the current prototype is still strongly dominated by copy operations through serialization, since the underlying Iox2 transport latency itself is largely independent of payload size \cite{iceoryx2_github}.
In addition, the measurements exhibit noticeable fluctuations, reflected in high maximum values and deviations between mean and median. The primary cause is likely CPU scheduling effects on the system. A reduction of these effects can be expected when using a real-time kernel combined with adjusted scheduling priorities.

Overall, the results indicate the functional feasibility of the prototype and provide initial evidence that the shared-memory-based approach reduces latencies compared to the original network-based implementation. The shared-memory-based implementation consistently achieves lower latencies than the original network-based variant despite the additional serialization overhead. However the achieved latencies are in the microsecond range which is generally expected, but appear relatively high compared to the transmission latencies of between 100\,-\,800\,ns \cite{iceoryx2_github} reported for Iox2. However, these values refer exclusively to pure data transfer, while this work relies on the Request/Response mechanism of the framework. Since, to the best of the authors knowledge, no benchmark for the reactivity of this mechanism is currently available, further investigation is required to assess the achievable performance limits.

\section{Conclusion and Future Work}
This paper presented a refined concept and prototypical implementation of an OPC UA based communication model for microservice-oriented automation systems, specialized for co-located applications. The proposed approach introduces a centralized modeling of semantic information on a Data Directory Server, enabling a unified representation and simplified discovery of available data. At the same time, data exchange between \emph{Providers} and \emph{Consumers} is defined as a peer-to-peer mechanism to reduce dependencies and minimize communication overhead.

To enable low-latency intra-host communication, the concept was implemented using the shared-memory-based middleware framework Iceoryx2. For this purpose, an existing OPC UA library was modified and evaluated in a proof-of-concept setup. The results show consistently lower latencies compared to the original network-based implementation, with improvements of approximately 15\,µs across different payload sizes, corresponding to 50\,\% for small payloads and around 30\,\% for larger ones in the best-case scenario. However, the current prototype remains influenced by serialization and copy operations, leading to higher latencies and a noticeable dependency on payload size, indicating further optimization potential.

Future work will therefore address the influence of payload size, with the goal of achieving a more constant latency behavior. Building on this, a more comprehensive evaluation of service reactivity will be conducted, including detailed benchmarks to quantify the gap to achievable lower bounds. In parallel, the central \emph{DataDirectoryServer} concept will be validated in a microservice architecture compared to the classical OPC UA Client/Server approach, with respect to scalability and system complexity.

\section*{Acknowledgements}

\noindent This work was conducted as part of the project "Modulare Automatisierung mit Mensch-Bot-Orchestrierung" (Modular automation with human-bot orchestration), MAMBO, which was co-financed by the European Regional Development Fund (ERDF/EFRE) and the German state of Rhineland-Palatinate.

\begingroup
\setlength{\itemsep}{0pt}

{\footnotesize                  % Schriftgröße Literatur kleiner
\bibliography{references}
\bibliographystyle{unsrtnat}
}

\endgroup

\end{document}